\documentclass[preprint2]{aastex}
\usepackage{graphicx}
\usepackage{color}

\begin{document}

\title{On the possibility of the Dyson spheres observable beyond the infrared spectrum}

\author{Osmanov Z. \& Berezhiani V. I.}
\affil{School of Physics, Free University of Tbilisi, 0183, Tbilisi,
Georgia}

\begin{abstract}
In this paper we revisit the Dysonian approach and assume that a superadvanced civilisation is capable of building a cosmic megastructure located closer than the habitable zone (HZ). Then such a Dyson Sphere (DS) might be visible in the optical spectrum. We have shown that for typical high melting point meta material - Graphene, the radius of the DS should be of the order of $10^{11}$cm, or even less. It has been estimated that energy required to maintain the cooling system inside the DS is much less than the luminosity of a star. By considering the stability problem, we have found that the radiation pressure might stabilise dynamics of the megastructure and as a result it will oscillate, leading to interesting observational features - anomalous variability. The similar variability will occur by means of the transverse waves propagating along the surface of the cosmic megastructure. In the summary we also discuss the possible generalisation of definition of HZs that might lead to very interesting observational features.

\end{abstract}

\keywords{Dyson sphere; SETI; Extraterrestrial; life-detection}

\section{Introduction}

A recent revival of interest to the search for advanced extraterrestrial civilisations 
has been provoked by the discovery of the object KIC8462852 observed by the 
Keppler mission \citep{kic846}. According to the data, the flux coming from the 
mentioned object has been characterised by dips of the order of $20\%$, which 
automatically excludes existence of a planet shading the star. It is worth noting 
that the detected flux has an irregular character, which has been analysed in detail 
and it has been shown that such a behaviour might not be caused by any measurement 
errors. Such a strange emission pattern of KIC8462852 and some other similar 
stars has been analysed in a series of papers in the context of artificial megastructures 
possibly constructed by a superadvanced civilisation \citep{wright,opt,radio}. 

The idea of such cosmic mega constructions has been considered by Freeman Dyson. 
In his well known work \citep{dyson} he has suggested that if the superadvanced 
extraterrestrials exist, which are capable of utilising the total energy of their host star (Level-II
 civilisation in the Kardashev scale \citep{kardashev}), then they 
could have built a relatively thin spherical shell (Dyson sphere (DS) with radius $\sim 1AU$ 
surrounding the star. The author has shown that on the mentioned distance the internal 
surface of the DS will be in the habitable zone (HZ). Therefore, the temperature of the 
megastructure will be of the order of $300$K, which means that it will be
visible in the infrared spectrum.

For finding the infrared DSs the monitoring of the sky has been performed by several 
instruments \citep{jugaku,slish,timofeev} although no real candidate was found. In the 
beginning of the last decade a new series of investigation has been initiated. In 
particular, \cite{carrigan} has examined and analysed the results of IRAS 
(The Infrared Astronomical Satellite) covering $96$\% of the sky and   
$16$ objects has been identified as potential DSs but it has been emphasised 
that the subsequent investigation is necessary. 

According to the Kardashev's classification the Level-III civilisation is the one that 
is capable of consuming the whole energy of its host galaxy by enveloping each of the 
star by spheres. This means that the galaxy will be seen in infrared. The search
for galaxy spanning civilisations has been performed by \cite{grif} where the authors 
analysed the data of the the Wide-field Infrared Survey Explorer (WISE) in the high
mid infrared spectrum. In total approximately $10^5$ galaxies have been monitored. 
They have identified almost $100$ objects which, according to the authors, deserve further study.

Last two decade researchers actively discuss the reasons of possible failure of 
conventional approach to the Search for Extraterrestrial Intelligence (SETI) \citep{cirkovic3,cirkovic1}.
\cite{cirkovic2} have considered that the so-called Dysonian method 
to SETI is not limited by the conventional approach and a wider 
view to the SETI problems is required. On the other
other hand, a good strategy for astronomy is to observe all interesting events in the sky 
in as many channels as possible. Recently a rather different view
to the Dysonian approach has been published. In particular, in \citep{paper1} the idea of 
freeman Dyson has been extended and the possibility of colonisation of a nearby region of a 
pulsar by a type-II civilisation was discussed. It has been shown that instead of a sphere the 
extraterrestrials should use a ring-like megaconstructions requiring much less material 
than for spheres. It has been found that locating in the HZ the rings will be 
visible in the infrared spectrum. In the following work \citep{paper2} the possibility of 
detection of infrared rings by modern facilities has been considered and it was found that in the nearby 
area of the solar system the monitoring of approximately $64\pm 21$ pulsars might be 
promising.

In the framework of the paradigm that conventional approaches should be widened,
in the present manuscript we examine an unusual view to the Dysonian SETI. In particular, 
if a civilisation has reached the level-II on the Kardashev's scale, then it might have been  
capable of living inside the shell of the DS even if it is not in the HZ of the star but 
closer to it. The smaller the radius of the DS the less the material it requires. 
therefore it is quite reasonable to search for such megastructures. On the other hand, the
closer the surface to the star the higher its temperature. Therefore, the observational
signature of this DS will be different and it might be visible in the optical spectrum, which according to our analysis
might exhibit anomalous variability. Relatively hotter DSs have been proposed by \cite{sandberg}, where the author considered the megastructures built for the reason of supercomputing.

In general, it is worth noting that in the current paper we consider existence of possibility of extraterrestrials
with biology similar to ours. But it is clear that this is a very limited approach, the so-called carbon chauvinism and water chauvinism ironically coined by Carl Sagan \citep{sagan}. In his book he examines the possibility of noncarbon and nonwater-based live that might significantly enrich a methodology of the search for extraterrestrials.

The organisation of the paper is the following: in Sec. 2 we present major
estimates and discuss possible observational patterns of the DSs and in Sec. 3 we 
summarise our results.

\section[]{Main consideration}
In this section we consider a theoretical model of DSs with reduced radii, estimate corresponding
temperatures and the consequent emission spectra and study the stability of the megastructure. 

Unlike the case examined by \cite{dyson}, where the DSs were located in the star's HZ in 
this section we consider a sphere with higher surface temperature, $T$. This could be quite 
reasonable, because smaller megastructure requires less material. By assuming  
the black body radiation one can show that the radius of the DS is given by
$$R =
\left(\frac{L}{4\pi\sigma T^4}\right)^{1/2}\approx $$
\begin{equation}
\label{R} 
\approx2.14\times 10^{12}\times
\left(\frac{L}{L_{\odot}}\right)^{1/2}\times
\left(\frac{1000K}{T}\right)^2 cm,
\end{equation}
where $\sigma\approx 5.67\times 10^{-5}$erg/(cm$^2$K$^4$) is the
Stefan-Boltzmann's constant and $L_{\odot}\approx 3.83\times 10^{33}$ergs s$^{-1}$ is the solar
luminosity. As it is clear from the above estimate the radius is almost one orders of magnitude less than one astronomical unit, a typical size of the DS in the HZ.  Here we used two assumptions: (I) the surface temperature is less than the melting point of a material the DS is made of and (II) the super civilisation is capable of constructing an efficient cooling system inside the shell. Generally speaking, it is clear that a Level-II civilisation might use meta materials with high melting temperatures. In particular, by considering graphene it is straightforward to show that  to maintain a cold region with a temperature of the order of $300$K the corresponding heat flux power (normalised on the solar luminosity) is given as
$$P_c\approx \frac{\kappa S}{L_{\odot}}\frac{\Delta T}{h}\approx 2.4\times10^{-6}\times$$
\begin{equation}
\label{thermal} 
\times \frac{\Delta T/h}{70K\; cm^{-1}}\times\frac{\kappa}{2.5\times 10^8 erg/(cmK)}\times\frac{S}{S_E}ergs \;s^{-1},
\end{equation}
where the temperature gradient is calculated for $\Delta T = (1000-300)K=700K$, $h = 100$cm and $\kappa=2.5\times 10^8$ergs cm$^{-1}$K$^{-1}$ is the typical value of the thermal conductivity of graphene\citep{graphene1} and $S$ - the area occupied by the extraterrestrials is normalised by the total surface area of Earth $S_E\approx 5.1\times 10^8$km$^2$. If one assumes that the coefficient of performance (COP) for a cooling system is of the order of $5$ (typical values of modern refrigerators), the engine, to compensate the aforementioned flux to the cold area must process the energy from the cold reservoir to the hot one

$$P_e=\frac{P_c}{COP}\approx 4.7\times10^{-7}\times\frac{5}{COP}\times $$
\begin{equation}
\label{thermal} 
\times\frac{\Delta T/h}{70K\; cm^{-1}}\times\frac{\kappa}{2.5\times 10^8 erg/(cmK)}\times\frac{S}{S_E}ergs \;s^{-1}.
\end{equation}
As we see from this estimate, only a tiny fraction of the total luminosity is required to maintain a habitable zone inside the DS if one uses meta materials (graphene), which Type-I civilisation can produce. Therefore, there is no question if the Level-II extraterrestrials can produce such materials. One has to note that although temperature dependence for graphene is not studied for a wide range of temperatures, a preliminary study shows that for high temperatures $\kappa$ should decrease \citep{graphene2}. Therefore, the corresponding value of $P_e$ might be even less. It is worth noting that if the primary purpose to construct the DS is computation, the aforementioned estimates should be changed \citep{sandberg}.

An important issue we would like to address is the stability problem of the DSs. In an ideal case the star has to be located in the centre of the sphere, but it is clear that a physical system can remain in the equilibrium state only in the absence of disturbances. In Fig. 1 we schematically show the position of the star (point S) and the DS
with the centre O. By assuming that the DS is shifted by $x$ with respect to the equilibrium position we intend to study its dynamics of motion. It is obvious that according to the Gauss' law the gravitational interaction of the star and the spherical shell is zero. On the other hand, stars emit enormous energy in the form of electromagnetic radiation, which will inevitably act on the internal surface of the DS and consequently, the part of the surface which is closer to the star will experience higher pressure than the opposite side of the sphere. Therefore the restoring force will appear leading to the periodic motion of the megastructure.

By assuming the isotropic radiation, the corresponding pressure in the direction of emission (See the arrow
in Fig.1) is given by
\begin{equation}
\label{press} 
P=\frac{L}{4\pi r^2c},
\end{equation}
where 
\begin{equation}
\label{r} 
r = \left(x^2+R^2-2xR\cos\varphi\right)^{1/2}
\end{equation}
is the distance from the centre of the radiation source.

If one assumes that the inner surface of the DS completely absorbs the incident radiation, the corresponding 
force acting on the differential surface area $dA$ is given by
\begin{equation}
\label{dF} 
dF = \frac{L}{4\pi r^2c}dA\cos\gamma,
\end{equation}
where $\gamma$ is the angle $\angle OBS$. On the other hand, from the symmetry it is certain that the sphere will move along the $x$ axis (coincident with the line $OS$). Therefore, the  dynamics is defined by the component of the force along the mentioned direction. Then, by considering a ring with radius $2\pi R\sin\varphi$ it is straightforward to show that 
\begin{equation}
\label{dFx} 
dF_x = \frac{L}{4\pi r^2c}2\pi R^2\sin\varphi\cos\gamma\cos\theta d\varphi.
\end{equation}

After taking into account the relations 
\begin{equation}
\label{theta} 
\theta = \varphi+\gamma,
\end{equation}

and 

\begin{equation}
\label{cosg} 
\cos\gamma = \frac{R-x\cos\varphi}{\left(x^2+R^2-2xR\cos\varphi\right)^{1/2}}
\end{equation}
the $x$ component of the total radiation force acting on the DS 
\begin{equation}
\label{Ftot} 
F_x = \frac{LR^2}{2c}\int_0^{\pi}\frac{\sin\varphi\left(R\cos\varphi-x\right)\left(R-x\cos\varphi\right)}
{\left(x^2+R^2-2xR\cos\varphi\right)^{2}}d\varphi,
\end{equation}
for small oscillations $x/R<< 1$ reduces to
\begin{equation}
\label{Ftot} 
F_x = -\frac{4L}{3c}\frac{x}{R}.
\end{equation}
The minus sign clearly indicates that the force has a restoring character and consequently the dynamics of the 
DS is described by the following differential equation
\begin{equation}
\label{eq} 
\frac{dx^2}{dt^2}+\frac{4L}{3cRM}x = 0,
\end{equation}
where $M=4\pi\rho R^2\Delta R$ is the mass of the megastructure, $\Delta R$ is its thickness and $\rho$ is 
the density of a material the DS is made of. By combining this expression with the aforementioned equation one can show that the period of oscillation is given by
$$P_{_{DS}} =2\pi\left(\frac{3\pi\rho cR^3\Delta R}{L}\right)^{1/2}\approx 33.8\times 
\left(\frac{L}{L_{\odot}}\right)^{1/4}\times$$
\begin{equation}
\label{per} 
\times \left(\frac{1000K}{T}\right)^{3/4}\times\left(\frac{\rho}{0.4g/cm^3}\right)^{1/2}
\times\left(\frac{\Delta R}{100cm}\right)^{1/2} yrs,
\end{equation}
where the thickness of the construction is normalised on $100$cm and $\rho$ is normalised on the density of graphene. As it is clear from Eq. (\ref{per}), the DS is stable oscillating with the period $\sim 33.8$ yrs. One can straightforwardly show that for the mentioned temperature the required mass of a megastructure is less than the mass of Earth. On the other hand, as we have already discussed, since smaller DSs will require less material, the issue that the civilisation might address would be to increase the surface temperature of the sphere. 

In Fig. \ref{fig2} we show the behaviour of $P_{_{DS}}$ versus the surface temperature for graphene having the melting temperature, $4510$K \citep{graphene}. Here we assume that if our civilisation (almost Level-I) might produce such materials, a super advanced one is able to do more efficiently. It is clear from the plot, that the period varies from $33.8$yrs to $0.53$yrs. This result automatically means that the search for Dysonian megastructures could be widened and observations should be performed not only in the infrared spectrum, but in the optical band. One of the characteristic features, that might distinguish the real DS from stars is the luminosity temperature relation. For example, main sequence stars having temperature of the order of $2000$K belong to low luminosity M stars \citep{carroll}, whereas the megastructure might emit the luminosity equal or even higher than the Solar luminosity. The temperature range we consider is $(2000K; 4000K)$, which means that the spectral radiance peaks at wavelengths $(725nm; 1450nm)$, having significant fraction in the optical band. Here we examined graphene as a particular example just to show that if extraterrestrials consider strong materials, their DS might exhibit an interesting behaviour different from normal stars.

Another important observational fingerprint follows from the results obtained above. Due to the oscillation of the hot DS its detected flux will be variable, with the period $P_{DS}$. From the figure it is clear that the values of $P_{_{DS}}$ are typical timescales of variability of very long period pulsating stars, which normally belong to spectral class: F, M, S or C. But F stars usually have very high temperatures, of the order of $7000$K, typical temperatures of M stars although are relatively low $(2400-3700)$K but their luminosities are almost two orders of magnitude less than the Solar luminosity. Unlike them, S and C stars are highly luminous (in comparison with the Solar luminosity) \citep{carroll}. Therefore, these discrepancies might be good indicators of potentially interesting objects. It is worth noting that the aforementioned examples only show a certain tendency how an observational pattern of the megastructures could be significantly different from the behaviour of variable stars. This in turn, means that an analysis of rich observational data provided by several optical telescopes might be very promising.

The surface of the DS is not an absolutely rigid body and therefore, it might vibrate under the influence 
of perturbations transversal to the surface, which might be induced either by the 
radiation pressure or by means of the star's wind. Therefore, another possible source of variability of DSs could be the mechanical waves generated on the $2D$ surface of the megaconstruction. It is straightforward to show that if a membrane is stressed with a tension per unit length, (along the surface of the DS) $\tau$, and its mass per unit area is $\Sigma$, the transverse wave speed along the surface is of the order of
\begin{equation}
\label{speed} 
\upsilon_w\approx\left(\frac{\tau}{\Sigma}\right)^{1/2}.
\end{equation}
Even if the construction does not envelope the star completely (Dyson Swarm) it still can have similar interesting observational features. In particular, by means of these waves the surface will vibrate exhibiting the variability of emission intensity. Unlike the completely closed surface, the Dyson swarm will experience the gravitational force from the Star leading to the following value of $\tau$
\begin{equation}
\label{force} 
\tau\approx \alpha G\frac{MM_s}{2\pi R^3},
\end{equation}
where $G\approx 6.67\times 10^{-8}$Nm$^2$/kg$^2$ is the gravitational constant, $M_s$ is the star's mass and $\alpha$ is the coefficient, that depends how complete the DS is (for the closed surface $\alpha=0$). We assumed that the tension is caused by gravitation (although the possible rotation of the DS might also influence the value of $\tau$). It is clear that the timescale, $P_w$, for waves to travel from one point to a diametrically opposite location is given by $\upsilon_w/(\pi R)$ which for Graphene will be $1.5\alpha$ yrs for $2000$K and $0.8\alpha$ yrs for $4000$K. In case of higher harmonics the corresponding values will be even less. Although we do not know the value of $\alpha$ and the origin of the tension is unclear, the estimate shows that an aforementioned anomalous variability might be a good sign for a potential DS.

\section{Conclusion}

We have generalised the Dysonian approach, considering megastructures not in the HZ but closer.
In the framework of the paper we assume that a super civilisation is capable of building megastructures
with melting point more than $2000K-4000K$. This means that the DSs having the length scales of the order $10^{11}$cm should be visible in the optical band as well.

It has been argued that the radiation pressure stabilises the DS, which potentially can lead to anomalous variability. In particular, by examining the super strong and super light material graphene and assuming that since a civilisation like us can produce it, the Level-II might have created even stronger and lighter material, it has been found that the variability would have been characterised by the timescales incompatible with known long period variable stars.

The similar variability might be caused by the transverse waves on the surface of the DS, where for an incomplete megastructure it has been shown that for super strong materials the "pulsation" period might be of the same order as in the aforementioned cases.

By the present paper we wanted to show that the Dysonian approach is broader and there are more possibilities of the search for intelligent life than it is sometimes thought. In this paper we have theoretically hypothesised our approach. 

Though the following generalisations are beyond the scope of the present paper, it is worth noting them. One of the significant issues one has to address is the question concerning anomalous variable intensity and the possibility to detect them by existent instruments.

Another issue that one should address is a certain extension of the Dysonian SETI in the context of HZs. In particular, as we have already discussed in the introduction, our approaches are  very restricted. Usually the HZ is defined as an area where water can be maintained in a liquid phase, whereas if one assumes nonwater-based life \citep{sagan} the corresponding area will be in a different location. According to Carl Sagan our chemistry is attuned to the temperature of our planet and he assumes that other temperatures might lead to other biochemistries. If this is the case the HZ will be an area where temperature might be significantly different from the temperature of liquid water. For instance, if one assumes that methane is used as a solvent instead of water, then the HZ will be an area where there are appropriate conditions to maintain methane in a liquid phase. In this case the temperature is in the range $90.7K-11.7K$ corresponding to the wavelength interval $26\mu m-32\mu m$. Therefore, in the framework of this paradigm, the DS in the HZ might be visible in the far IR spectrum.


\begin{figure}
  \centering {\includegraphics[width=7cm]{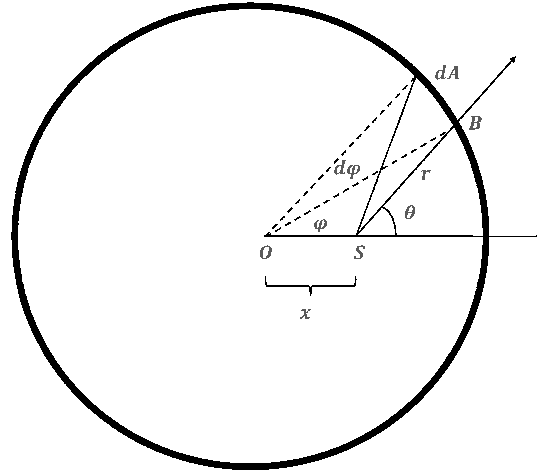}}
  \caption{Here we schematically show the DS with the center O and a star S inside it.}\label{fig1}
\end{figure}

\begin{figure}
  \centering {\includegraphics[width=7cm]{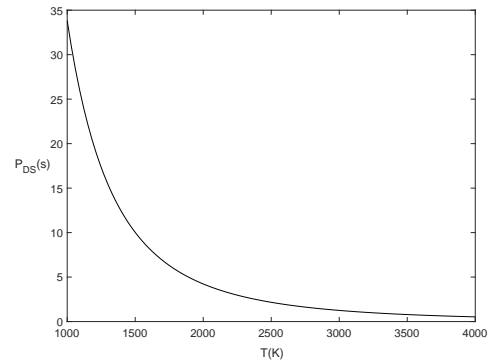}}
  \caption{Here we plot the dependence of oscillation period of the DS for 3D Graphene with the density $\rho = 0.4$g/cm$^{3}$ and the melting point, $T_m \approx 4510$K.}\label{fig2}
\end{figure}

\section*{Acknowledgments}
The research was supported by the Shota Rustaveli National Science Foundation grant (DI-2016-14). 


\begin{thebibliography}{99}
\bibitem[Boyajian et al.(2016)]{kic846} Boyajian, T.S. et al., 2016,
MNRAS, 457, 3988 
\bibitem[Bradbury et al.(2011)]{cirkovic1} Bradbury, R. J., Cirkovic, M. M. \& Dvorsky, G.,
2011, JBIS, 64, 156
\bibitem[Cai et al.(2010)]{graphene1} Cai, W. et al., 2010, Nano Letters, 10,1645
\bibitem[Carrigan(2009)]{carrigan} Carrigan, R. A., 2009, ApJ, 698, 2075 
\bibitem[Carroll \& Ostlie(2010)]{carroll} Carroll, Bradley W. \& Ostlie, Dale A., 2010, An introduction to modern astrophysics and cosmology. Pearson
\bibitem[Cirkovic(2008)]{cirkovic3} Cirkovic, M. M., 2008,
AsBio, 4, 225
\bibitem[Cirkovic \& Bradbury(2006)]{cirkovic2} Cirkovic, M. M. \& Bradbury, R. J., 2006,
New Astr., 11, 628


\bibitem[Dyson(1960)]{dyson} Dyson, F., 1960,
Science, 131, 1667 
\bibitem[Griffith et al.(2015)]{grif} Griffith, R. L., Wright, J.T., Maldonaldo, J., Povich, M. S.,
Sigurdsson, S. \& Mullan, B, 2015, ApJ, 217, 25

\bibitem[Harp et al.(2016)]{radio} Harp, G. R., Richards, J., Shostak, S.,
Tarter, J. C., Vakoch, D. A. \& Munson, C., 2016, ApJ, 825, 1 

\bibitem[Jugaku \& Nishimura(2002)]{jugaku} Jugaku, J. \&
Nishimura, S., 2002, in Proc. IAU Symp. 213, Bioastronomy 2002: Life
Among the Stars, ed. R. Norris \& F. Stootman (San Francisco, CA:
ASP), 437 
\bibitem[Kardashev(1964)]{kardashev} Kardashev, N. S., 1964,
AJ, 8, 217 
\bibitem[Los et al.(2015)]{graphene} Los, J. H., Zakharchenko, K. V., Katsnelson, M. I.
\& Fasolino, Annalisa, 2015, PhRev B, 91, 045415


\bibitem[Osmanov(2017)]{paper2} Osmanov, Z., 2017, IJAsB, (Published online, 
doi:10.1017/S1473550417000155)
\bibitem[Osmanov(2016)]{paper1} Osmanov, Z., 2016, IJAsB, 15, 127

\bibitem[Pop et al.(2012)]{graphene2} Pop, E., Varshney, V. \& Roy, A., 2012, MRSBulletin, 37, 1273
\bibitem[Sagan(2000)]{sagan} Sagan, C., 2000, The Cosmic Connection: 
An Extraterrestrial Perspective, ed. Jerome Agel
\bibitem[Sandberg(1999)]{sandberg}  Sandberg, A., 1999, J. Evol. Tech., 5,1
\bibitem[Schuetz et al.(2016)]{opt} Schuetz, M., Vakoch, D. A., Shostak, S.
\& Richards, J., 2016, ApJL, 825, L5 
\bibitem[Slish(1985)]{slish} Slish, V. I., 1985, in The Search for
Extraterrestrial Life: Recent Developments, ed. M. D. Papagiannis
(Boston, MA: Reidel Pub. Co.), 315 
\bibitem[Timofeev et al.(2000)]{timofeev} Timofeev, M. Y., Kardashev, N. S. \& Promyslov, V. G.,
2000, Acta Astronautica J., 46, 655 
\bibitem[Wright et al.(2016)]{wright} Wright, J. T., Cartie, K. M., Kimberly M. S., Zhao, M.,
Jontof-Hutter, D. \& Ford, E. B., 2016, ApJ, 816, 22 
\end{thebibliography}
\end{document}